# Voltage Control of a van der Waals Spin-Filter Magnetic Tunnel Junction

Tiancheng Song,[1] Matisse Wei-Yuan Tu,[2] Caitlin Carnahan,[3] Xinghan Cai,[1] Takashi Taniguchi,[4] Kenji Watanabe,[4] Michael A. McGuire,[5] David H. Cobden,[1] Di Xiao,[3] Wang Yao,[2] Xiaodong Xu[1,6]*

[1]Department of Physics, University of Washington, Seattle, Washington 98195, USA.
[2]Department of Physics and Center of Theoretical and Computational Physics, University of Hong Kong, Hong Kong, China.
[3]Department of Physics, Carnegie Mellon University, Pittsburgh, Pennsylvania 15213, USA.
[4]National Institute for Materials Science, Tsukuba, Ibaraki 305-0044, Japan.
[5]Materials Science and Technology Division, Oak Ridge National Laboratory, Oak Ridge, Tennessee 37831, USA.
[6]Department of Materials Science and Engineering, University of Washington, Seattle, Washington 98195, USA.

*Correspondence to: xuxd@uw.edu

**Abstract:**

**Atomically thin chromium triiodide ($CrI_3$) has recently been identified as a layered antiferromagnetic insulator, in which adjacent ferromagnetic monolayers are antiferromagnetically coupled[1,2]. This unusual magnetic structure naturally comprises a series of anti-aligned spin filters which can be utilized to make spin-filter magnetic tunnel junctions with very large tunneling magnetoresistance (TMR)[3–6]. Here we report voltage control of TMR formed by four-layer $CrI_3$ sandwiched by monolayer graphene contacts in a dual-gated structure. By varying the gate voltages at fixed magnetic field, the device can be switched reversibly between bistable magnetic states with the same net magnetization but drastically different resistance (by a factor of ten or more). In addition, without switching the state, the TMR can be continuously modulated between 17,000% and 57,000%, due to the combination of spin-dependent tunnel barrier with changing carrier distributions in the graphene contacts. Our work demonstrates new kinds of magnetically moderated transistor action and opens up possibilities for voltage-controlled van der Waals spintronic devices.**

**Main Text:**

Electrical manipulation of magnetism is central to spintronics[7–13]. Voltage-controlled switching between bistable magnetic states can be employed in energy efficient magnetic memory and logic technologies. In this regard, the recently discovered two-dimensional (2D) magnetic insulator chromium triiodide ($CrI_3$) has several assets as a building block for van der Waals (vdW) spintronics[1,2,14]. First, the extreme thinness of few-layer $CrI_3$ enhances the probability that the magnetism will be amenable to electrostatic control[15–20]. Second, the layered antiferromagnetic structure at zero field naturally forms a series of interlayer spin filters, and their relative alignment can be changed by a moderate magnetic field via spin-flip transitions. This unusual property underpins the recent demonstration of multiple-spin-filter magnetic tunnel junctions (sf-MTJs) that exhibit giant tunneling magnetoresistance (TMR)[3–6,21,22].

In multilayer CrI$_3$, for a given net magnetization there are multiple nearly degenerate magnetic states with different patterns of layer magnetization[3]. Switching between these states reconfigures the interlayer spin filters and thus can change the tunneling resistance. If this switching could be induced by voltage alone it would represent a new kind of magnetic logic. We explore this possibility using a sf-MTJ with four-layer CrI$_3$ tunnel barrier between monolayer graphene contacts, as shown schematically in Fig. 1a. This sf-MTJ is sandwiched between two hexagonal boron nitride (hBN) flakes with a graphite top gate (held at voltage $V_{tg}$) and SiO$_2$/Si substrate used as a bottom gate (at voltage $V_{bg}$). The monolayer graphene contacts combine a low density of states with high carrier mobility[23,24], allowing much stronger gating effects than using conventional metal electrodes in a vertical junction structure. The tunneling current ($I_t$) is measured while applying a DC bias voltage ($V$) to the top graphene contact with the bottom one grounded. All measurements described in the main text, except where specified, were made on the device whose optical image is shown in Fig. 1b (device 1), at a temperature of 2 K.

We first use reflective magnetic circular dichroism (RMCD) to probe the net magnetization[3]. Figure 1c shows the RMCD signal as a function of out-of-plane magnetic field ($\mu_0 H$) swept from negative to positive (orange curve) and vice versa (green curve). It displays typical four-layer CrI$_3$ behavior[3]. At low fields (<0.7 T) the net magnetization nearly vanishes, corresponding to either of the two fully antiferromagnetic states, ↑↓↑↓ or ↓↑↓↑, as indicated in the figure. The arrows here denote the out-of-plane magnetization from top to bottom layer respectively. The small remnant

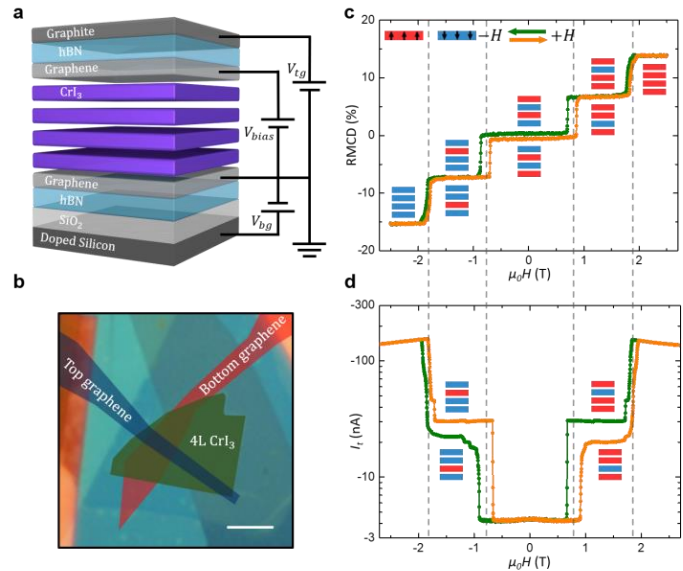

**Figure 1 | Magnetic states in four-layer CrI$_3$ spin-filter magnetic tunnel junction (sf-MTJ).** **a**, Schematic of a four-layer CrI$_3$ sf-MTJ device including two monolayer graphene contacts and top and bottom gates. **b**, False-color optical micrograph of device 1 (scale bar 5 µm). **c**, Reflective magnetic circular dichroism (RMCD) signal as a function of out-of-plane magnetic field ($\mu_0 H$) from device 1. The orange (green) curve corresponds to sweeping the magnetic field up (down). Insets show the corresponding magnetic states. The red and blue blocks denote the out-of-plane magnetization of individual layers pointing up and down, respectively. **d**, Tunneling current ($I_t$) of the same device at representative bias and gate voltages ($V$ = -240 mV, $V_{tg}$ = 0 V and $V_{bg}$ = 0 V).

RCMD signal is caused by a slight asymmetry between the top and bottom layers due to the fabrication process[3]. At high fields (>2 T) the RMCD signal saturates, corresponding to the fully aligned magnetic states, ↑↑↑↑ and ↓↓↓↓. Since the behavior is essentially the same for the opposite field direction, from now on in the discussion we focus on positive magnetic fields.

At intermediate fields (between 0.9 T and 1.7 T) the RMCD signal is about half the saturated value, implying that the net magnetization is half that of the fully aligned state. This is consistent with any of the set of four magnetic states where one layer has the opposite magnetization to the other three, {↑↓↑↑, ↑↑↓↑, ↓↑↑↑, ↑↑↑↓}. Among these, the first two have two antiparallel interfaces while the last two have only one such interface. Since antiparallel interfaces are favored by the antiferromagnetic coupling, the first two should have lower energy. Therefore, we expect the magnetic configuration at intermediate fields to be either ↑↓↑↑ or ↑↑↓↑, these being degenerate and almost indistinguishable if the two internal layers are equivalent.

If a bias is applied either across the junction or between the gates, the RMCD, and thus the net magnetization, does not change, but the electric field can lift the symmetry between ↑↓↑↑ and ↑↑↓↑ and the two states may thus respond differently to the bias, which is expected to yield distinct tunneling magnetoresistance. Figure 1d shows $I_t$ as a function of $\mu_0 H$ with a bias $V = -240$ mV on the top graphene and both gates grounded. By comparing with Fig. 1b, we see that the lowest and highest current plateaus correspond to the antiferromagnetic and fully aligned magnetic states, respectively. Interestingly, at intermediate magnetic fields where there is only one plateau in the RMCD signal there are two distinct plateaus in the tunneling current. These must correspond to ↑↓↑↑ and ↑↑↓↑, that is, the current is sensitive to which of the internal layers has the minority magnetization. Modeling the system as a set of coupled magnetic quantum wells, we find that ↑↓↑↑ carries the higher tunneling current than ↑↑↓↑ under these conditions, as the former has a transmission resonance closer to the bias window (supplementary materials). We also conclude that the system is bistable, remaining in either one of these two magnetic configurations if the field is kept in the intermediate range.

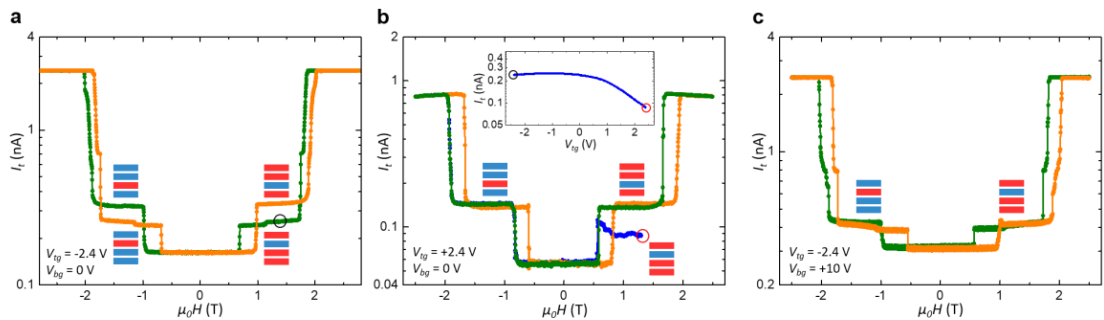

**Figure 2 | Electric control of bistable magnetic states. a**, **b**, **c**, $I_t$ ($V = +80$ mV) as a function of $\mu_0 H$ at three representative gate voltages with identified magnetic states as shown in the insets. The orange (green) curve corresponds to sweeping the magnetic field up (down). The inset of **b** shows $I_t$ as sweeping $V_{tg}$ from -2.4 to +2.4 V. The black and red open circles indicate the starting and end points, corresponding to the circles in **a** (initial state) and **b** (final state). Subsequently, $\mu_0 H$ sweeps down and $I_t$ is monitored, as shown by the blue curve in the main panel of **b**.

Now we turn to our key finding, which is that switching between the bistable magnetic states can be controlled and induced by gate voltage, and that this affects the tunneling current. Figures 2a-c show $I_t$ (at $V = +80$ mV) as the field is swept up and down at three selected pairs of gate voltages. At $V_{tg} = -2.4$ V, $V_{bg} = 0$ V (Fig. 2a) we see two intermediate-field plateaus, as in Fig. 1d, implying that ↑↓↑↑ and ↑↑↓↑ are similarly stable. However, at $V_{tg} = +2.4$ V, $V_{bg} = 0$ V (Fig. 2b, orange and green curves) we see only a higher current plateau, while at $V_{tg} = -2.4$ V, $V_{bg} = +10$ V (Fig. 2c) we see only a lower current plateau. This suggests that the latter gate-voltage pairs cause either ↑↑↓↑ or ↑↓↑↑ to be preferred, respectively.

To confirm this, we first prepare the system in the low-current state at 1.3 T with the gate voltage pair set at the bistable condition of Fig. 2a, indicated by the black open circle in Fig. 2a. We then sweep $V_{tg}$ from -2.4 to +2.4 V, finishing in the gate voltage condition of Fig. 2b. While doing this we monitor the current, which decreases smoothly (inset to Fig. 2b) to the level at the point indicated by the red open circle in Fig. 2b. When the magnetic field is subsequently swept down (blue curve in Fig. 2b) the current jumps to a lower value below 0.7 T, and thereafter repeated cycling between ±2.5 T simply reproduces the prior behavior with a single intermediate plateau. From these observations we infer that if the system is prepared in the state ↑↓↑↑ at $V_{tg} = -2.4$ V, then sweeping $V_{tg}$ with the magnetic field fixed is an adiabatic process that maintains it in this state. However, at $V_{tg} = +2.4$ V this state is only metastable, and it cannot be entered from either fully aligned (highest current) or antiferromagnetic (lowest current) states merely by sweeping the magnetic field.

Remarkably, at larger bias voltages it is possible to induce reversible switching between the two magnetic states purely by gate voltage control, while staying at a single fixed magnetic field. Figures 3a and c show the current at $V = -240$ mV vs magnetic field at $V_{tg} = -2.4$ V and +2.4 V, respectively, both at $V_{bg} = 0$ V. The inferred magnetic states are indicated by insets; note that the state that carries the higher current at this negative bias is the one that carried the lower current at the positive bias $V = +80$ mV (see Fig. 2a). This is self-consistent with bistable states assignment, since the reversal of current flow direction accompanies the reversal of the relative magnitude of $I_t$ between the bistable magnetic states.

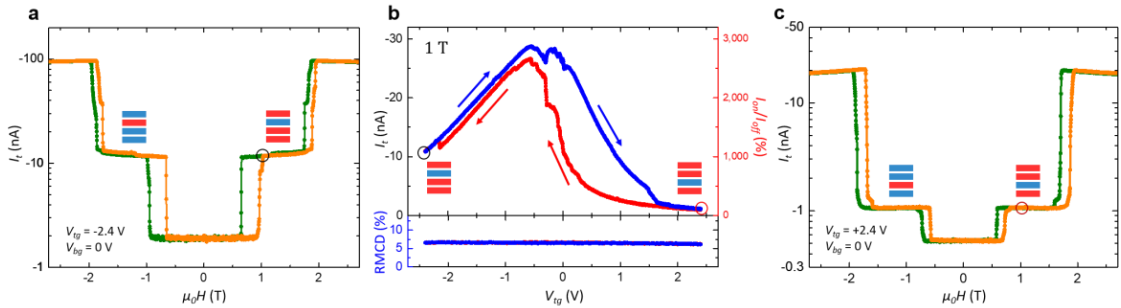

**Figure 3 | Reversible voltage switching of the bistable magnetic states. a**, **c**, $I_t$ ($V = -240$ mV) as a function of $\mu_0 H$ at two representative gate voltages with identified magnetic states shown in the insets. **b**, $I_t$ and the extracted magnetoresistance ratio as a function of $V_{tg}$ swept from +2.4 V to -2.4 V (red curve) and back to +2.4 V (blue curve) at fixed $\mu_0 H = 1$ T. The black and red open circles denote the two ends of the voltage sweep loop, corresponding to the same states circled in **a** and **c**, respectively. The hysteresis curve demonstrates magneto-electric coupling. The bottom panel shows little changes in RMCD during the voltage sweep, consistent with the equal magnetization of the bistable states.

If we now fix $\mu_0 H = 1$ T and sweep $V_{tg}$ up and down between -2.4 V and +2.4 V (Fig. 3b), the current changes repeatably between end values corresponding to ↑↓↑↑ and ↑↑↓↑ (determined from Figs. 3a & c), implying that reproducible switching between these states occurs. Meanwhile, the RMCD signal is almost constant (bottom panel, Fig. 3b), as expected since the two states have the same net magnetization. The general changes in the current with increasing magnitude of $V_{tg}$ is probably associated with doping of the graphene contacts causing a different mismatch of spin or momentum between the contacts. Most interestingly, at intermediate $V_{tg}$ there is pronounced hysteresis in the current, just as expected for a transition between two metastable states, accompanied by small wiggles that are naturally explained by associated domain effects. Within this hysteretic region the current differs between the two states by as much as a factor of ten.

The influence of the gate voltages on the magnetic states can be in principle due to modifications of the anisotropy and interlayer coupling through changes in orbital occupancy and/or electric-field effects modifying the energy splitting of ↑↓↑↑ and ↑↑↓↑. Monte Carlo simulations (supplementary materials) reveal that changing anisotropy alone is not sufficient, and changing interlayer coupling must be included to reproduce the experimental observation.

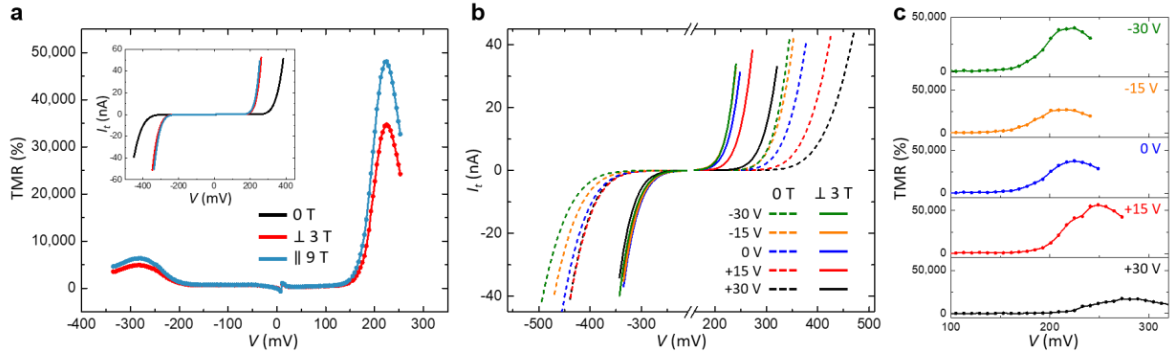

**Figure 4 | Gate tunable tunneling magnetoresistance. a**, TMR ratio as a function of $V$ derived from the $I_t$-$V$ data shown in the inset, at $V_{bg} = 0$ V. **b**, $I_t$-$V$ curves at five representative gate voltages for layered-antiferromagnetic ground states (dashed curves, at 0 T) and fully aligned magnetic states (solid curves, at 3 T). **c**, TMR ratio as a function of $V$ at a series of $V_{bg}$.

Finally, we show that direct and dramatic gate-voltage control of the TMR is possible in such devices, irrespective of the switching effect described above. As usual, we define the TMR ratio by $100\% \times (R_{ap}-R_p)/R_p$, where $R_{ap}$ and $R_p$ are the DC resistances with fully anti-parallel (antiferromagnetic, low-field) and parallel (fully spin-polarized, high-field) layer magnetization measured at a given bias. Figure 4a shows the TMR as a function of bias for device 2 (which has a single bottom gate), derived from $I_t$-$V$ curves as shown in the inset for both in-plane and out-of-plane magnetic field at $V_{bg} = 0$ V. The behavior is similar to that reported previously for ungated devices[3]. The TMR is substantially modified by back gate voltage. Figure 4b shows $I_t$-$V$ curves for layered-antiferromagnetic states (dashed curves, $\mu_0H = 0$ T) and fully aligned states (solid curves, $\mu_0H = 3$ T), at $V_{bg}$ values between -30 V and +30 V. There is a consistent shift of the thresholds in these curves towards positive bias as $V_{bg}$ becomes more positive. Figure 4c shows the derived TMR ratio. Its peak value varies from 57,000% to 17,000%. The origin of this behavior is under investigation and beyond the scope of this work, but it may involve a combination of electric field modification of the spin-dependent tunnel barrier (supplementary materials)[25,26], changes of Fermi level and magnetic proximity effects induced by $CrI_3$ in the monolayer graphene contacts[27,28].

**References:**

1. Huang, B. *et al.* Layer-dependent ferromagnetism in a van der Waals crystal down to the monolayer limit. *Nature* **546**, 270–273 (2017).

2. Seyler, K. L. *et al.* Ligand-field helical luminescence in a 2D ferromagnetic insulator. *Nat. Phys.* **14**, 277–281 (2018).

3. Song, T. *et al.* Giant tunneling magnetoresistance in spin-filter van der Waals heterostructures. *Science* **360**, 1214–1218 (2018).

4. Klein, D. R. *et al.* Probing magnetism in 2D van der Waals crystalline insulators via electron tunneling. *Science* **360**, 1218–1222 (2018).

5. Wang, Z. *et al.* Very large tunneling magnetoresistance in layered magnetic semiconductor $CrI_3$. *Nat. Commun.* **9**, 2516 (2018).

6. Kim, H. H. *et al.* One million percent tunnel magnetoresistance in a magnetic van der Waals heterostructure. Preprint at https://arxiv.org/abs/1804.00028 (2018).

7. Matsukura, F., Tokura, Y. & Ohno, H. Control of magnetism by electric fields. *Nat. Nanotech.* **10**, 209–220 (2015).

8. Ohno, H. *et al.* Electric-field control of ferromagnetism. *Nature* **408**, 944–946 (2000).

9. Chiba, D., Yamanouchi, M., Matsukura, F. & Ohno, H. Electrical manipulation of magnetization reversal in a ferromagnetic semiconductor. *Science* **301**, 943–945 (2003).

10. Weisheit, M. *et al.* Electric field-induced modification of magnetism in thin-film ferromagnets. *Science* **315**, 349–351 (2007).

11. Maruyama, T. *et al.* Large voltage-induced magnetic anisotropy change in a few atomic layers of iron. *Nat. Nanotech.* **4**, 158–161 (2009).

12. Wang, W.-G., Li, M., Hageman, S. & Chien, C. L. Electric-field-assisted switching in magnetic tunnel junctions. *Nature Mater.* **11**, 64–68 (2012).

13. Shiota, Y. *et al.* Induction of coherent magnetization switching in a few atomic layers of FeCo using voltage pulses. *Nat. Mater.* **11**, 39–43 (2011).

14. Huang, B. *et al.* Electrical control of 2D magnetism in bilayer $CrI_3$. *Nat. Nanotech.* https://doi.org/10.1038/s41565-018-0121-3 (2018).

15. Zhong, D. *et al.* Van der Waals engineering of ferromagnetic semiconductor heterostructures for spin and valleytronics. *Sci. Adv.* **3**, e1603113 (2017).

16. Geim, A. K. & Grigorieva, I. V. Van der Waals heterostructures. *Nature* **499**, 419–425 (2013).

17. Novoselov, K. S., Mishchenko, A., Carvalho, A. & Castro Neto, A. H. 2D materials and van der Waals heterostructures. *Science* **353**, aac9439 (2016).

18. Jiang, S., Shan, J. & Mak, K. F. Electric-field switching of two-dimensional van der Waals magnets. *Nat. Mater.* http://doi.org/10.1038/s41563-018-0040-6 (2018).

19. Jiang, S., Li, L., Wang, Z., Mak, K. F. & Shan, J. Controlling magnetism in 2D $CrI_3$ by electrostatic doping. https://doi.org/10.1038/s41565-018-0135-x (2018).

20. Sivadas, N., Okamoto, S. & Xiao, D. Gate-controllable magneto-optic Kerr effect in layered collinear antiferromagnets. *Phys. Rev. Lett.* **117,** 267203 (2016).

21. Worledge, D. C. & Geballe, T. H. Magnetoresistive double spin filter tunnel junction. *J. Appl. Phys.* **88**, 5277–5279 (2000).

22. Miao, G.-X., Müller, M. & Moodera, J. S. Magnetoresistance in double spin filter tunnel junctions with nonmagnetic electrodes and its unconventional bias dependence. *Phys. Rev. Lett.* **102**, 076601 (2009).

23. Novoselov, K. S. *et al.* Two-dimensional gas of massless Dirac fermions in graphene. *Nature* **438**, 197–200 (2005).

24. Zhang, Y., Tan, Y.-W., Stormer, H. L. & Kim, P. Experimental observation of the quantum Hall effect and Berry's phase in graphene. *Nature* **438**, 201–204 (2005).

25. Lezlinger, M. & Snow, E. H. Fowler-Nordheim tunneling into thermally grown $SiO_2$. *J. Appl. Phys.* **40**, 278–283 (1969).

26. Britnell, L. *et al.* Field-effect tunneling transistor based on vertical graphene heterostructures. *Science* **335**, 947–950 (2012).

27. Greenaway, M. T. *et al.* Resonant tunnelling between the chiral Landau states of twisted graphene lattices. *Nat. Phys.* **11**, 1057–1062 (2015).

28. Wei, P. *et al.* Strong interfacial exchange field in the graphene/EuS heterostructure. *Nat. Mater.* **15**, 711–716 (2016).

29. Blake, P. *et al.* Making graphene visible. *Appl. Phys. Lett.* **91**, 063124 (2007).

30. Zomer, P. J., Guimarães, M. H. D., Brant, J. C., Tombros, N. & van Wees, B. J. Fast pick up technique for high quality heterostructures of bilayer graphene and hexagonal boron nitride. *Appl. Phys. Lett.* **105**, 013101 (2014).

31. Sato, K. Measurement of magneto-optical Kerr effect using piezo-birefringent modulator. *Jpn. J. Appl. Phys.* **20**, 2403–2409 (1981).

## Methods:

### Device fabrication

CrI$_3$ crystals were mechanically exfoliated onto 90 nm SiO$_2$/Si substrates in a nitrogen glove box with water and oxygen concentration less than 0.5 ppm. The four-layer CrI$_3$ flakes were identified by their optical contrast relative to the substrate using the established optical contrast models of CrI$_3$[1,29]. The monolayer graphene, graphite and 5-30 nm hBN flakes were exfoliated onto either 285 nm or 90 nm SiO$_2$/Si substrates and examined by optical and atomic force microscopy under ambient conditions. Only atomically clean and smooth flakes were identified and used. Metallic V/Au (7/70 nm) electrodes were deposited onto the bottom hBN flakes and substrates using electron beam evaporation before a standard electron beam lithography with a bilayer resist (A4 495 and A4 950 poly (methyl methacrylate (PMMA))). The van der Waals stacking was performed in the glove box using a polymer-based dry transfer technique[30]. The flakes were picked up sequentially: top gate graphite, top hBN, top monolayer graphene contact, four-layer CrI$_3$, bottom monolayer graphene contact. The resulting stacks were then transferred and released on top of the bottom hBN with pre-patterned electrodes. In the complete heterostructure, the CrI$_3$ flake is fully encapsulated, and the top/bottom monolayer graphene and the top gate graphite flakes are connected to the pre-patterned electrodes.

### Electrical measurement

The electrical measurements were performed in a PPMS DynaCool cryostat (Quantum Design, Inc.) with a base temperature of 1.7 K. The four-layer CrI$_3$ sf-MTJ devices were mounted in a Horizontal Rotator probe, which allows applying out-of-plane or in-plane magnetic field up to 9 T. Figure 1a shows the schematic of four-layer CrI$_3$ sf-MTJs. The DC bias voltage ($V$) is applied to the top monolayer graphene contact with the bottom monolayer graphene contact grounded. The top and bottom gate voltages ($V_{tg}$ and $V_{bg}$) are applied to the top gate graphite and bottom doped Si substrate, respectively. The resulting tunneling current ($I_t$) is amplified and measured by a current preamplifier (DL Instruments; Model 1211).

### Reflective magnetic circular dichroism measurement

The reflective magnetic circular dichroism (RMCD) measurements were performed in an attocube closed-cycle cryostat (attoDRY 2100) with a base temperature of 1.55 K and up to 9 T magnetic field in the out-of-plane direction. A power-stabilized 632.8 nm HeNe laser was used to probe the device at normal incidence with a fixed power of 1 μW. The AC lock-in measurement technique used to measure the RMCD signal follows closely to the previous magneto-optical Kerr effect (MOKE) and RMCD measurements of the magnetic order in atomically-thin CrI$_3$[1,3,31].

**Acknowledgements:** We thank Yongtao Cui for insightful discussion. Funding: Work at the University of Washington was mainly supported by the Department of Energy, Basic Energy Sciences, Materials Sciences and Engineering Division (DE-SC0018171). Device fabrication and part of transport measurements are supported by NSF-DMR-1708419, NSF MRSEC 1719797, and UW Innovation Award. D.H.C. is supported by DE-SC0002197. Work at CMU is supported by DOE BES DE-SC0012509. Work at HKU is supported by the Croucher Foundation (Croucher Innovation Award), RGC of HKSAR (17303518P). Work at ORNL (M.A.M.) was supported by the US Department of Energy, Office of Science, Basic Energy Sciences, Materials Sciences and Engineering Division. K.W. and T.T. acknowledge support from the Elemental Strategy Initiative conducted by the MEXT, Japan and JSPS KAKENHI Grant Numbers JP15K21722. D.X. acknowledges the support of a Cottrell Scholar Award. X.X. acknowledges the support from the State of Washington funded Clean Energy Institute and from the Boeing Distinguished Professorship in Physics.

**Author Contributions:** X.X. and T.S. conceived the project. T.S. fabricated the devices, performed the experiments, and analyzed the data, assisted by X.C., supervised by X.X., D.X., W.Y., and D.H.C.. M.W.-Y.T. and W.Y. modeled the tunneling current. C.C. and D.X. performed the Monte Carlo simulation. M.A.M. provided and characterized bulk $CrI_3$ crystals. T.T. and K.W. provided and characterized bulk hBN crystals. T.S., X.X., D.H.C., D.X., and W.Y. wrote the manuscript with input from all authors.

**Competing Financial Interests:** The authors declare no competing financial interests.

**Data Availability**: The data that support the findings of this study are available from the corresponding author upon reasonable request.

# Supplementary Information for

# Voltage Control of a van der Waals Spin-Filter Magnetic Tunnel Junction

Tiancheng Song,[1] Matisse Wei-Yuan Tu,[2] Caitlin Carnahan,[3] Xinghan Cai,[1] Takashi Taniguchi,[4] Kenji Watanabe,[4] Michael A. McGuire,[5] David H. Cobden,[1] Di Xiao,[3] Wang Yao,[2] Xiaodong Xu[1,6]*

[1]Department of Physics, University of Washington, Seattle, Washington 98195, USA.
[2]Department of Physics and Center of Theoretical and Computational Physics, University of Hong Kong, Hong Kong, China.
[3]Department of Physics, Carnegie Mellon University, Pittsburgh, Pennsylvania 15213, USA.
[4]National Institute for Materials Science, Tsukuba, Ibaraki 305-0044, Japan.
[5]Materials Science and Technology Division, Oak Ridge National Laboratory, Oak Ridge, Tennessee 37831, USA.
[6]Department of Materials Science and Engineering, University of Washington, Seattle, Washington 98195, USA.

*Correspondence to: xuxd@uw.edu

**Contents:**

1. **Measurements on an additional four-layer CrI$_3$ sf-MTJ device**
2. **Configuration differentiated tunneling currents**
3. **Monte Carlo simulation**

# S1. Measurements on an additional four-layer CrI$_3$ sf-MTJ device

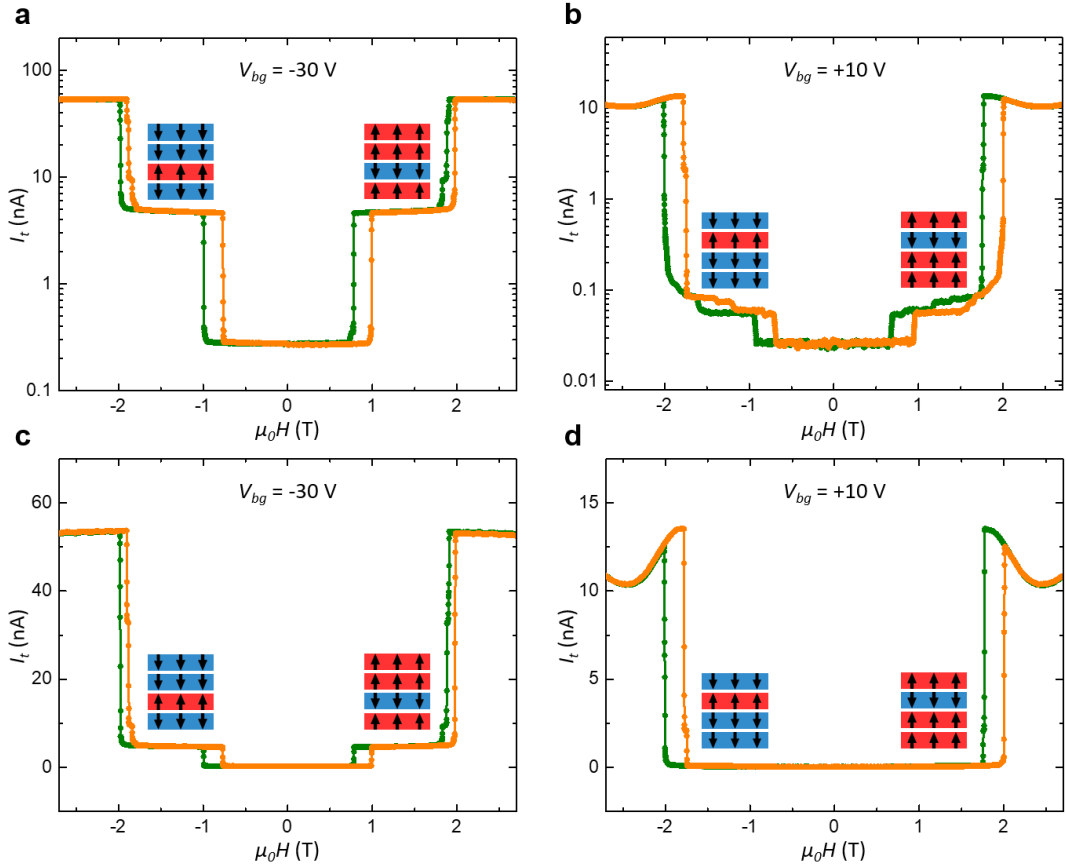

**Figure S1 | Electric control of the magnetic states in device 2. a**, **b**, $I_t$ ($V = +250$ mV) as a function of $\mu_0 H$ at two representative gate voltages plotted on a semi-log scale, with identified magnetic states shown in the insets. **c**, **d**, The same data in **a** and **b** replotted on a linear scale. Device 2 has a single bottom gate, and $V$ is applied to the top graphene contact with the bottom one grounded.

## S2. Configuration differentiated tunneling currents

Here we provide an intuitive explanation for why the two 3:1 configurations, namely, ↑↓↑↑ and ↑↑↓↑, pass different currents for the same bias.

### S2.1 Determining which 3:1 configuration yields larger tunneling current

The experimentally observed current-voltage curves show onsets of rapidly increasing current at certain applied biases. This indicates that the transport bias window (formed by the difference between the two chemical potentials of the contacts) lies below the conduction bands in the CrI$_3$ such that it is the low-energy tail of the transmission, rather than the resonant part of it, that is responsible for the tunneling current (so as to give rise to an onset behavior). Therefore, the configuration whose transmission spectrum has a stronger tail in the low-lying transport window is the one that produces larger current.

To determine which configuration has a lower energy transmission tail, we make the following observations. (i) The spin-down part of the bands lies higher above the spin-up part by the amount of the spin-splitting energy which is the largest energy scale. It is larger than the applied bias. Since the transport window lies lower than the spin-up bands, tunneling from the contacts to the spin-down bands is relatively weak in comparison to the spin-up bands. We therefore only need to consider the currents contributed by the spin-up component. (ii) Due to the van der Waals gaps, the inter-layer hopping as well as tunneling to the contacts is small in comparison to the spin-splitting as well as the applied bias and the in-plane dispersion. Henceforth, the relevant transmission can be attributed to contributions from each layer and they mainly peak around the respective band energies. (iii) The applied bias shifts the bands between the layers uniformly with distance along the direction of the electric field. The relevant spin-up transmission profiles are schematically plotted for the two configurations respectively as Fig. S2a and Fig. S2b.

We assume that the in-plane momentum conservation occurs only between the four layers of the tunneling junction. The Hamiltonian for the few-layer tunnel junction under a given in-plane momentum is $H_{TJ} = \sum_{\sigma=\uparrow,\downarrow} H_\sigma$, with $H_\sigma = \sum_{n=1}^{N_l} \varepsilon_{\sigma n} d_{\sigma n}^+ d_{\sigma n} + t\left(\sum_{n=1}^{N_l-1} d_{\sigma n}^+ d_{\sigma n+1} + h.c.\right)$, describing the energy $\varepsilon_{\sigma n}$ for layer $n$ and spin $\sigma$ with tunneling between neighboring layers with hopping $t$. Here $N_l = 4$ is the number of layers and other details can be found in S2.2. The lowest eigenenergy of the spin-up component of the tunnel junction Hamiltonian, $H_\uparrow$, thus gives a good indication about where the lowest resonance of transmission lies. By comparing the lowest eigenenergies of $H_\uparrow$ for the two configurations, one can henceforth infer which configuration has lower transmission tail and thus resulting in larger current.

For clarity we call $\uparrow\downarrow\uparrow\uparrow$ and $\uparrow\uparrow\downarrow\uparrow$ the configurations $c1$ and $c2$ respectively and the corresponding Hamiltonian $H_\uparrow^{c1}$ and $H_\uparrow^{c2}$ are explicitly given by,

$$H_\uparrow^{c1} = \begin{pmatrix} -\frac{\Delta_0}{2} + 3\delta\varepsilon + \varepsilon_\parallel & t & 0 & 0 \\ t & +\frac{\Delta_0}{2} + 2\delta\varepsilon + \varepsilon_\parallel & t & 0 \\ 0 & t & -\frac{\Delta_0}{2} + \delta\varepsilon + \varepsilon_\parallel & t \\ 0 & 0 & t & -\frac{\Delta_0}{2} + \varepsilon_\parallel \end{pmatrix},$$

$$H_\uparrow^{c2} = \begin{pmatrix} -\frac{\Delta_0}{2} + 3\delta\varepsilon + \varepsilon_\parallel & t & 0 & 0 \\ t & -\frac{\Delta_0}{2} + 2\delta\varepsilon + \varepsilon_\parallel & t & 0 \\ 0 & t & +\frac{\Delta_0}{2} + \delta\varepsilon + \varepsilon_\parallel & t \\ 0 & 0 & t & -\frac{\Delta_0}{2} + \varepsilon_\parallel \end{pmatrix}.$$

Here $\Delta_0$ denotes the spin-splitting, $\delta\varepsilon = \frac{V}{N_l-1}$ in which $V$ is the applied bias with $0 < V < \Delta_0$ and $\varepsilon_\parallel$ is the in-plane kinetic energy. By treating the hopping as a perturbation taking into account the lowest non-vanishing correction, the lowest eigenenergies of $H_\uparrow^{c1}$ and $H_\uparrow^{c2}$ are found to be respectively

$$E_\uparrow^{c1} = \varepsilon_\parallel - \frac{\Delta_0}{2} - \frac{t^2}{\delta\varepsilon}, \qquad E_\uparrow^{c2} = \varepsilon_\parallel - \frac{\Delta_0}{2} - \frac{t^2}{\Delta_0 + \delta\varepsilon}.$$

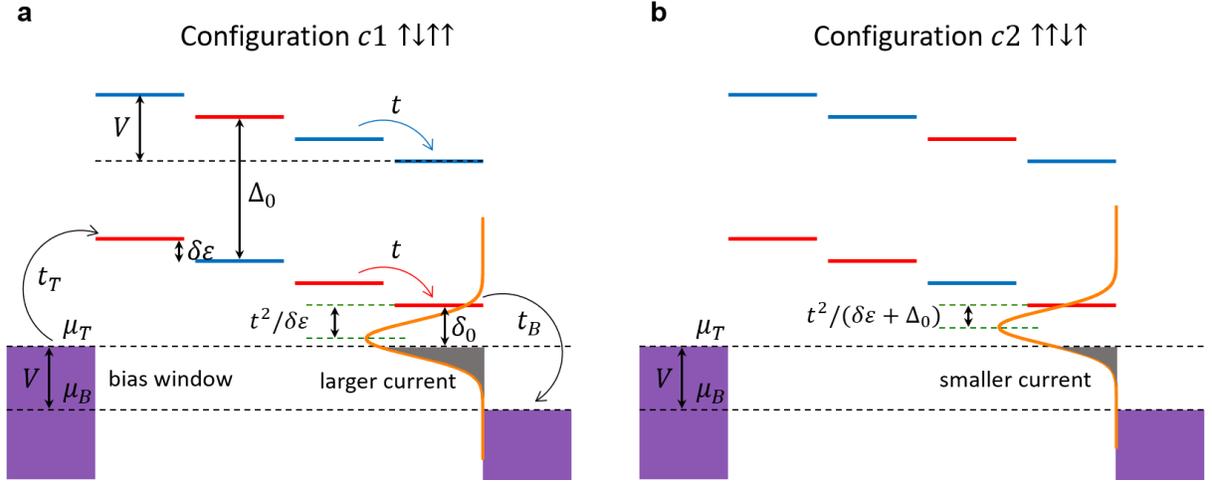

**Figure S2 | Schematics of resonant levels in a four-layer CI$_3$ tunnel junction distinguishing tunneling current. a, b,** The configurations for ↑↓↑↑ and ↑↑↓↑. The blue solid lines and the red solid lines indicate the energy levels for spin-down and spin-up respectively. Due to inter-layer hopping, the resonance in transmission is shifted away from native position (red and blue solid lines). The profile of transmission is schematically sketched as the orange line shape whose overlap with the bias window (gray shaded area) determines the amount of current that goes through the tunnel junction. The difference between the amount of transmission resonance's shift from the original place in the two configurations makes one be favored than the other in transporting current.

Since $\frac{t^2}{\delta\varepsilon} > \frac{t^2}{\Delta_0+\delta\varepsilon} > 0$, clearly $E_\uparrow^{c1}$ lies lower in energy than $E_\uparrow^{c2}$. Therefore, one expects that the current obtained at configuration $c1$ is larger than that at $c2$. Note that reversing the bias reverses the arrangement of the energy levels. This results in smaller down-shift of the lowest eigenenergy of $c1$ in comparison to that of $c2$. Which configuration yields larger current is thus reversed by reversing the applied bias, consistent with the experimental observation.

This simple analysis is supported by full calculation of the currents using the Wingreen-Meir formula, also taking into account the energy-spreading of the levels due to in-plane momentum. The calculated current from spin-up component is exemplified by $\frac{I^{c1}}{I^{c2}} = 1.3$ with the parameters: $\Delta_0 = 0.5$ eV, $V = \mu_T - \mu_B = 0.08$ eV with $\Gamma_{\alpha\sigma} = t = 0.005$ eV and $\mu_T = -\frac{\Delta_0}{2} - \delta_0$ in which $\delta_0 = 0.07$ eV and $W = 0.6\,\Delta_0$.

### S2. 2 Details of the model for calculating the currents

Following the usual convention, we have the Hamiltonian for the few-layer tunnel junction in contacts with the electrodes given by, $H = H_{TJ} + H_E + H_T$, in which the first term has been given in S2.1. The second term is for the two contacts, $H_E = \sum_\sigma \sum_{\alpha=T,B} \varepsilon_{\sigma\alpha k} c^+_{\sigma\alpha k} c_{\sigma\alpha k}$, where $\alpha = T, B$ correspond to the top and bottom contacts. The last term accounts for the tunneling between the central layers and the contacts, $H_T = \sum_\sigma \sum_{\alpha=T,B} t_{\alpha\sigma} c^+_{\sigma\alpha k} d_{\sigma n} + h.c.$ Here the operator $d^+_{\sigma n}(d_{\sigma n})$ creates (annihilates) an electron with spin $\sigma$, in the $n$th layer with energy $\varepsilon_{\sigma n}$. Similarly, $c^+_{\sigma\alpha k}(c_{\sigma\alpha k})$ creates (annihilates) an electron with energy $\varepsilon_{\sigma\alpha k}$, spin $\sigma$ and momentum $k$ in contact $\alpha$ as an electron reservoir. The tunneling current from spin $\sigma$ is calculated using the standard Wingreen-Meir formula, $I = \int dE (f_T(E) - f_B(E)) T_\sigma(E)$, in which $f_\alpha(E)$ is the Fermi distribution function of contact $\alpha$ and $T_\sigma(E) = tr(\Gamma_{T\sigma} G^r_\sigma(E) \Gamma_{B\sigma} G^a_\sigma(E))$ is the transmission with $\Gamma_{\alpha\sigma} = 2\pi |t_{\alpha\sigma}|^2 \varrho$, where $\varrho$ is the density of states of the electron reservoirs and $G^{r/a}_\sigma(E)$ is the retarded/advanced Green function for electrons with spin $\sigma$. This current is further averaged over $\varepsilon_\parallel$ of a range between 0 and $W$ with a constant weighting for each energy since the density of states of a 2D electron gas is constant.

## S3. Monte Carlo Simulation

In our Monte Carlo simulation, we model the four-layer CrI3 as four antiferromagnetically coupled ferromagnetic layers, each composed of $40 \times 40$ spins on a square lattice. A single-spin Metropolis algorithm is used with the spin Hamiltonian given by

$$H = -J_{FM} \sum_{l,\langle i,j\rangle} \mathbf{S}_{l,i} \cdot \mathbf{S}_{l,j} + \sum_{l,i} J_{AFM,l} \mathbf{S}_{l,i} \cdot \mathbf{S}_{l+1,i} + K_l S^2_{l,i,z} - B S_{l,i,z},$$

where $\mathbf{S}_{l,i}$ is the spin unit vector residing on site $i$ of layer $l$. The $J_{FM}$ ($> 0$) term characterizes intralayer ferromagnetic exchange interactions and the $J_{AFM,l}$ ($> 0$) term characterizes antiferromagnetic coupling between layers $l$ and $l+1$. The $K_l$ term describes easy-axis anisotropy for $K_l < 0$ and the $B$ term describes Zeeman coupling. In our simulations, $J_{FM} = 1$ and magnetization per spin is calculated by averaging the $z$ components of all spins. When sweeping the magnetic field strength $B$, magnetization per spin is measured over the entire system as well as on each individual layer. Mean magnetization per spin measurements are calculated by averaging the mean $S_z$ value over 50,000 Monte Carlo steps at the indicated external field strength. We found that when sweeping the magnetic field strength in a system that exhibits inversion symmetry, namely, when all anisotropies and interlayer couplings are the same, respectively, all of the patterns of Fig. S3.1 appear with equal probability.

As mentioned in the main text, applying a gate voltage to the top or bottom of the system breaks the inversion symmetry. This effect is modeled in the simulation by adjusting the anisotropy $K$ and/or interlayer coupling strength $J_{AFM,l}$ and making them layer dependent. An example of this approach, where $K_1$ is increased from 0.05 to 0.08, is shown in Fig. S3.2a. By increasing the anisotropy on the top layer, we observe behavior similar to that shown in Fig. 2b in the main text. That is, when reversing the sweeping direction, the preferred bistable state in the intermediate magnetic field region is switched. We can obtain similar behavior by increasing $J_{AFM,1}$, the interlayer coupling between the top two layers. An example of this scenario is shown in Fig. S3.2b, where $J_{AFM,1}$ is increased from 0.1 to 0.12.

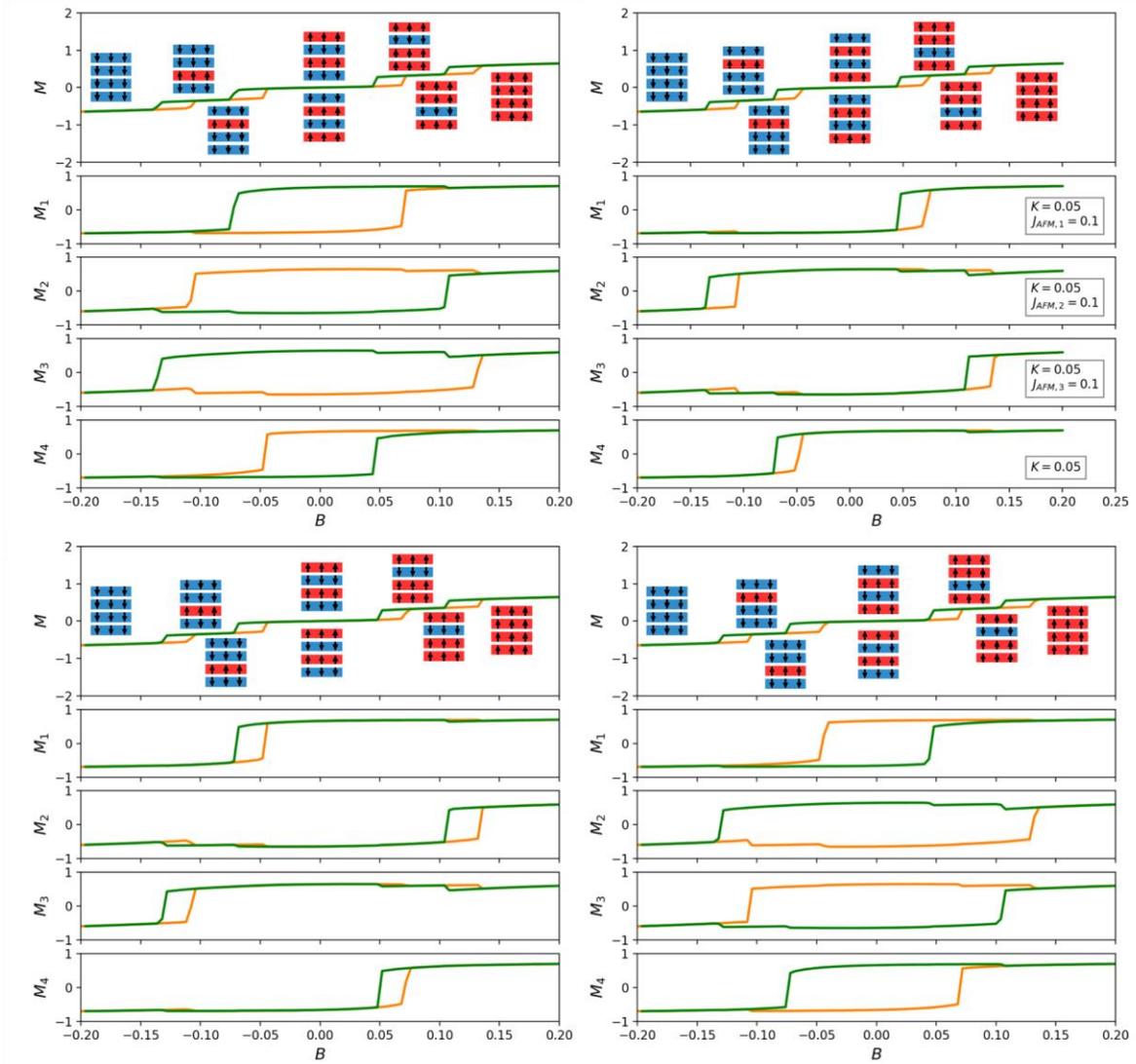

**Figure S3.1 | Magnetization versus magnetic field strength for the total system, *M*, and each of the individual layers, $M_i$.** Shown are the four possible switching patterns obtained for a symmetric system when sweeping the magnetic field. The orange line indicates sweeping of the magnetic field strength in the positive direction, and the green line indicates sweeping in the negative direction. All results presented are for four square lattice layers of $40 \times 40$ spins each at $T = 0.15$ with $K = 0.05$, $J_{AFM} = 0.1$, and $J_{FM} = 1$. The anisotropic and antiferromagnetic coupling strengths are shown in the inset of the individual layer plots, where $J_{AFM}$ displayed on the plot for layer $i$ refers to the coupling between layers $i$ and $i + 1$. Layer resolved magnetizations are also shown as insets in the plot for total magnetization $M$.

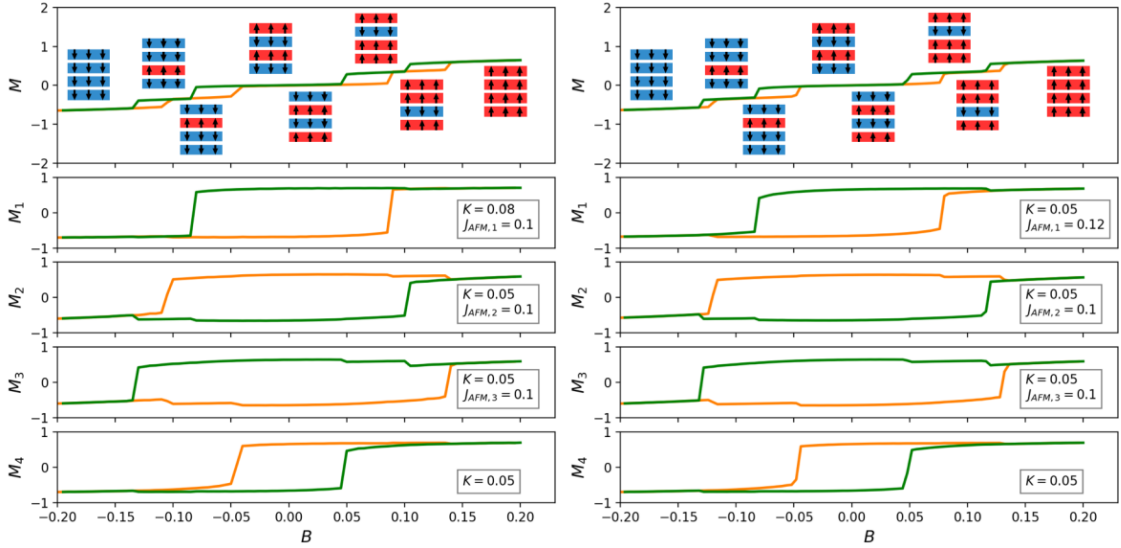

**Figure S3.2 | Magnetization versus magnetic field strength for systems with broken inversion symmetry. a**, The magnetic switching pattern observed when anisotropy is increased for the top layer only. In this case, $K_1 = 0.08$ while all other layers have an anisotropic strength of 0.05. **b**, The same magnetic switching pattern is observed by increasing the interlayer coupling strength between the top two layers. In this case, $J_{AFM,1} = 0.12$ while $J_{AFM} = 0.1$ throughout the rest of the system.

We now turn to a simulated demonstration of controlled, reversible switching between the bistable states. We utilize magnetic configurations that are independent of the external field sweeping direction, and thus model the character of the experimental approach described in Figs. 2 and 3 of the main text. While applying a gate voltage could change both the anisotropy and interlayer coupling, we find that in our simulation changing the anisotropy alone cannot reproduce the reversible state switching behavior observed experimentally. The dominant mechanism by which controlled and reversible bistable state switching may be obtained is the modification of the interlayer coupling strength.

We begin by referring to Fig S3.3a in which inversion symmetry is broken by increasing $J_{AFM,1}$, the interlayer coupling strength between layers 1 and 2. We note that the state ↑↓↑↑ is preferred in the intermediate field region where $B \cong 0.25$. This preference is independent of the sweeping direction and by fixing the external field around this value, we are guaranteed to observe this magnetic state. In Fig S3.3b, however, we see that we can invert the magnetic switching pattern by decreasing $J_{AFM,1}$. In this case, the state ↑↑↓↑ is always preferred in approximately the same intermediate field region.

By fixing the external field in this region, we may switch between preferred states by modifying $J_{AFM,1}$ as shown in Fig S3.4. In this case, the external field is fixed at $B = 0.23$, the anisotropy is set uniformly at $K = 0.05$, and the interlayer coupling is set at $J_{AFM} = 0.2$ between all layers except $J_{AFM,1}$ which is swept from 0.3 to 0.1 and back. The initial state of the system for $J_{AFM,1} = 0.3$ is ↑↓↑↑, consistent with the configuration observed in Fig S3.3a. As $J_{AFM,1}$ decreases to 0.1, the magnetizations of the middle two layers flip and we obtain the state ↑↑↓↑, consistent with Fig S3.3b. In sweeping $J_{AFM,1}$ back to 0.3, we recover the initial state ↑↓↑↑, demonstrating the reversibility of this approach for switching between the bistable states.

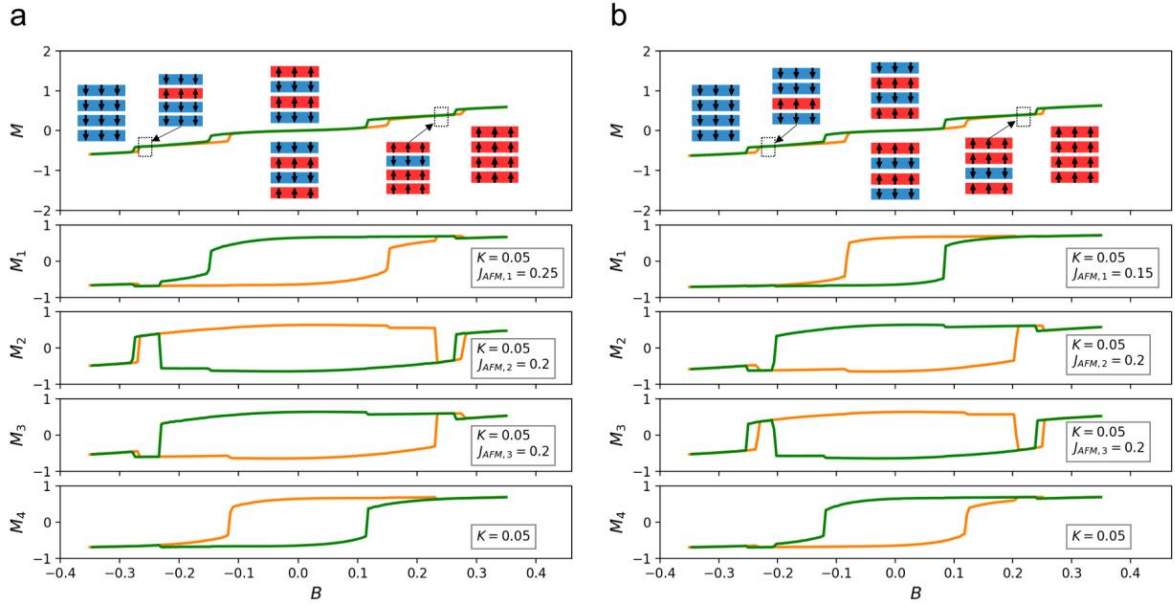

**Figure S3.3 | Broken inversion symmetry via interlayer coupling modification. a**, The magnetic switching pattern observed when interlayer coupling is increased between the top two layers only. In this case, $J_{AFM,1} = 0.25$ while all other layers have a coupling strength of 0.2. **b**, The same scenario with the top interlayer coupling decreased to $J_{AFM,1} = 0.15$. In both cases, we observe that a single magnetic configuration is preferred, regardless of sweeping direction, within a narrow intermediate field region.

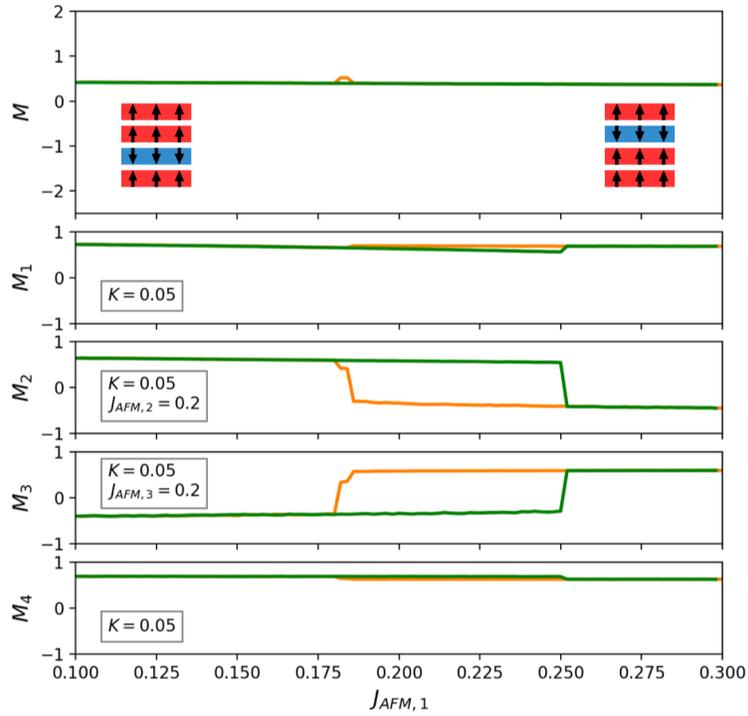

**Figure S3.4 | Bistable state switching via $J_{AFM,1}$ modification.** The external field is fixed at $B = 0.23$ while the interlayer coupling between the top two layers, $J_{AFM,1}$, is swept between 0.3 and 0.1. The initial state is ↑↓↑↑ at $J_{AFM,1} = 0.3$. Sweeping $J_{AFM,1}$ from 0.3 to 0.1 (green line) causes the system to transition into the state ↑↑↓↑. Sweeping back from 0.1 to 0.3 (orange line) restores the system to the initial state.

9